\input jytex.tex   
\typesize=10pt
\magnification=1200
\baselineskip17truept
\hsize=6truein\vsize=8.5truein
\sectionnumstyle{blank}
\chapternumstyle{blank}
\chapternum=1
\sectionnum=1
\pagenum=0

\def\begintitle{\pagenumstyle{blank}\parindent=0pt\begin{narrow}[0.4in]}
\def\endtitle{\end{narrow}\newpage\pagenumstyle{arabic}}


\def\beginexercise{\vskip 20truept\parindent=0pt\begin{narrow}[10 
truept]}
\def\endexercise{\vskip 10truept\end{narrow}}


\def\eql#1{\eqno\eqnlabel{#1}}
\def\ref{\reference}
\def\peq{\puteqn}
\def\pref{\putref}

\def\mgn{\marginnote}
\def\bex{\begin{exercise}}
\def\eex{\end{exercise}}


\def\mbox#1{{\leavevmode\hbox{#1}}}

\def\hspace#1{{\phantom{\mbox#1}}}

\def\rS{{\rm S}}

\def\al{\alpha}
\def\bGa{{\bmit\Gamma}} 
\def\bxi{{\bmit\xi}}

\def\ga{\gamma}
\def\de{\delta}
\def\Ga{\Gamma}

\def\ka{\kappa}
\def\la{\lambda}
\def\La{\Lambda}

\def\Om{\Omega}

\def\si{\sigma}

\def\th{\theta}

\def\De{\Delta}
\def\cau{{\cal U}}

\def\zf{$\zeta$--function}


\def\frac#1/#2{\leavevmode\kern.1em
\raise.5ex\hbox{\the\scriptfont0 #1}\kern-.1em/\kern-.15em
\lower.25ex\hbox{\the\scriptfont0 #2}}
\def\sfrac#1/#2{\leavevmode\kern.1em
\raise.5ex\hbox{\the\scriptscriptfont0 #1}\kern-.1em/\kern-.15em
\lower.25ex\hbox{\the\scriptscriptfont0 #2}}

\def\gtorder{\mathrel{\raise.3ex\hbox{$>$}\mkern-14mu
             \lower0.6ex\hbox{$\sim$}}}
\def\ltorder{\mathrel{\raise.3ex\hbox{$<$}\mkern-14mu
             \lower0.6ex\hbox{$\sim$}}}

\def\semidirprod{\rlap{\ss C}\raise1pt\hbox{$\mkern.75mu\times$}}
\def\for{\lower6pt\hbox{$\Big|$}}
\def\fish{\kern-.25em{\phantom{abcde}\over \phantom{abcde}}\kern-.25em}


\def\boxit#1{\vbox{\hrule\hbox{\vrule\kern3pt
        \vbox{\kern3pt#1\kern3pt}\kern3pt\vrule}\hrule}}
\def\dalemb#1#2{{\vbox{\hrule height .#2pt
        \hbox{\vrule width.#2pt height#1pt \kern#1pt
                \vrule width.#2pt}
        \hrule height.#2pt}}}

\def\frac#1#2{{{#1}\over{#2}}}


\def\eg{{\it e.g. }}
\def\ie{{\it i.e. }}

\def\pa{\partial}


\def\Tr{{\rm Tr\,}}

\def\3j#1#2#3#4#5#6{\left\lgroup\matrix{#1&#2&#3\cr#4&#5&#6\cr}
\right\rgroup}

\def\man{{\cal M}}

\def\m?{\mgn{?}}

\def\pa{\partial}

\def\beq{\begin{eqnarray}}
\def\eeq{\end{eqnarray}}


\def\aop#1#2#3{{\it Ann. Phys.} {\bf {#1}} (19{#2}) #3}

\def\cmp#1#2#3{{\it Comm. Math. Phys.} {\bf {#1}} (19{#2}) #3}
\def\cqg#1#2#3{{\it Class. Quant. Grav.} {\bf {#1}} (19{#2}) #3}

\def\jgp#1#2#3{{\it J. Geom. and Phys.} {\bf {#1}} (19{#2}) #3}
\def\jmp#1#2#3{{\it J. Math. Phys.} {\bf {#1}} (19{#2}) #3}
\def\jpa#1#2#3{{\it J. Phys.} {\bf A{#1}} (19{#2}) #3}

\def\np#1#2#3{{\it Nucl. Phys.} {\bf B{#1}} (19{#2}) #3}
\def\pl#1#2#3{{\it Phys. Lett.} {\bf {#1}} (19{#2}) #3}

\def\prD#1#2#3{{\it Phys. Rev.} {\bf D{#1}} (19{#2}) #3}

\def\cras#1#2#3{{\it Comptes Rend. Acad. Sci. (Paris)} {\bf{#1}} (#2) #3}

\def\mpcps#1#2#3{{\it Math. Proc. Camb. Phil. Soc.} {\bf{#1}} (19{#2}) #3}

\def\am#1#2#3{{\it Acta Mathematica} {\bf {#1}} (19{#2}) #3}
\def\aim#1#2#3{{\it Adv. in Math.} {\bf {#1}} (19{#2}) #3}
\def\ajm#1#2#3{{\it Am. J. Math.} {\bf {#1}} ({#2}) #3}

\def\aom#1#2#3{{\it Ann. of Math.} {\bf {#1}} (19{#2}) #3}

\def\cpde#1#2#3{{\it Comm. Partial Diff. Equns.} {\bf {#1}} (19{#2}) #3}

\def\invm#1#2#3{{\it Invent. Math.} {\bf {#1}} (19{#2}) #3}
\def\ijpam#1#2#3{{\it Ind. J. Pure and Appl. Math.} {\bf {#1}} (19{#2}) #3}
\def\jdg#1#2#3{{\it J. Diff. Geom.} {\bf {#1}} (19{#2}) #3}

\def\jmpa#1#2#3{{\it J. Math. Pures. Appl.} {\bf {#1}} ({#2}) #3}

\def\ojm#1#2#3{{\it Osaka J.Math.} {\bf {#1}} ({#2}) #3}

\def\pja#1#2#3{{\it Proc. Jap. Acad.} {\bf {A#1}} (19{#2}) #3}

\def\tams#1#2#3{{\it Trans. Am. Math. Soc.} {\bf {#1}} (19{#2}) #3}

\begin{title}  
\vglue 1truein
\vskip15truept
\centertext {\Bigfonts \bf Heat-kernel coefficients for oblique}
\vskip3truept
\centertext{\Bigfonts \bf boundary conditions}
\vskip 20truept 
\centertext{J.S.Dowker\footnote{dowker@a3.ph.man.ac.uk},\quad
Klaus Kirsten\footnote{Klaus.Kirsten@itp.uni-leipzig.de}} 
\vskip 7truept
\centertext{*\it Department of Theoretical Physics,\\
The University of Manchester, Manchester, England}
\vskip10truept
\vskip10truept
\vskip7truept
\centertext{\dag\it Universit\"at Leipzig, Institut f\"ur Theoretische 
Physik,\\ Augustusplatz 10, 04109 Leipzig, Germany}
\vskip 20truept
\centertext {Abstract}
\vskip10truept
\begin{narrow}
We calculate the heat-kernel coefficients, up to $a_2$, for a U(1) bundle
on the 4-Ball for boundary conditions which are such that the normal
derivative of the field at the boundary is related to a first-order
operator in boundary derivatives acting on the field. The results are
used to place restrictions on the general forms of the coefficients. In
the specific case considered, there can be a breakdown of ellipticity.
\end{narrow}
\vskip 5truept
\vskip 60truept
\vfil
\end{title}
\pagenum=0

The coefficients in the small-time asymptotic expansion of the heat-kernel
for a second order operator $\De$ play important roles both in quantum field
theory and pure mathematics. For bounded manifolds, the simplest boundary 
conditions are those that make the operator self-adjoint, this
usually being the case when physics dictates the conditions. $\De$ is
typically the Laplacian, $\De_2$, on a Riemannian manifold, $\man$.

The traditional conditions are Dirichlet and Neumann, and also a 
generalisation of the latter, sometimes referred to as Robin conditions,
but which go back at least to Newton. Recently, in response to questions
arising in quantum gravity, gauge theory and string theory, 
[\pref{barv,McandO,AandE,MandS}],
a generalised form of Robin conditions that involves {\it tangential},
as well as normal derivatives has been found necessary. In the mathematical 
literature these conditions are sometimes referred to as {\it oblique}. More 
generally, the boundary condition becomes nonlocal on the boundary.
The operator $\De$ can remain self-adjoint.

Some of the lower  heat-kernel coefficients (up to $a_1$) have been
evaluated by McAvity and Osborn [\pref{McandO}] using their extension of 
the recursion method developed by De Witt, following Hadamard, for closed 
Riemannian manifolds. Another
approach has been expounded by Avramidi and Esposito [\pref{AandE}] 
based on the functorial
methods most systematically used by Branson and Gilkey [\pref{BandG2}]. They 
also computed the coefficients in the very special case of a flat ambient 
manifold with a totally geodesic, flat boundary.

Although these boundary conditions have been the subject of classical analysis
(see \eg[\pref{EandS,Gilkey1,krantz,treves}]), the explicit determination of 
the heat-kernel coefficients is in its infancy and
in this letter we wish to report on this question using the approach of 
special case evaluation as particularly described
in [\pref{BGKE}] on the ball. (See also [\pref{BKD}]). 
We are able to obtain more information on 
some higher coefficients that, in conjunction with the functorial technique, 
will aid in determining their general forms, should these be required.

The evaluation is somewhat more involved than that for the standard
conditions which are local on the boundary so we shall simply outline the
method, giving the results and some commentary just to show what can be 
achieved. It will not be possible to work in arbitrary dimensions,
which was a feature of [\pref{BKD}], rather we will be restricted to 
dimension four.Nevertheless, sufficient interesting information will 
emerge to justify exposure now.

A hidden element of the calculation is the free use of algebraic 
manipulation. Because of the complexity of the expressions, this is
almost necessary but should be viewed simply as convenient bookkeeping.


McAvity and Osborn [\pref{McandO}] have employed a generalised form of Robin 
conditions,
written in the way adopted by Avramidi and Esposito [\pref{AandE}]
$$
{\cal B}=\nabla_N+{i\over2}\big(\Ga^i\widehat\nabla_i+\widehat\nabla_i
\Ga^i\big)-S,
\eql{gbc}$$
which involves tangential (covariant) derivatives, $\widehat\nabla_i$, computed
from the induced metric on the boundary. 
$\Ga^i$ is a bundle endomorphism valued boundary vector field
satisfying $\Ga^\dagger=\Ga$ (our $\Ga_i$ differs from that in [\pref{AandE}]
by a factor of $i$) and $S$ is a hermitian bundle automorphism. $\nabla_N$ is, 
for us, the {\it outward} normal derivative at the boundary.

The condition on a section of some vector bundle $V$ is
$$
{\cal B}V\big|_{\pa\man}=0.
\eql{bcss}$$

The first two nontrivial heat-kernel coefficients for the differential operator
$-\De_2+E$ are written in the form used by
Avramidi and Esposito, 
$$\eqalign{
a_{1/2}(f)=&{\sqrt\pi\over2}\int_{\pa\man}f\Tr(\ga),\cr
a_1(f)=&{1\over6}\int_{\man}f\,\Tr\big(\al_1E+\al_2R\big)\cr
&+{1\over6}\int_{\pa\man}\Tr\big(f(b_0\ka+b_2S)+b_1f_N\big)\cr
&\hspace{***}+{1\over6}\int_{\pa\man}f\,\Tr\big(\si_1\ka_{ij}
\Ga^i\Ga^j\big),\cr}
\eql{lowc}$$
where $f$ is a test (`smearing') function, $f_N$ being its normal derivative.
The `numerical' coefficients computed by McAvity and Osborn are
$$\eqalign{
\ga&={2\over(1-\Ga^2)^{1/2}}-1\cr
b_0&=2-6\bigg({1\over2\Ga}\log{1+\Ga\over1-\Ga}-{1\over1-\Ga^2}\bigg)\cr
b_1&=3\bigg(1-{1\over\Ga}\log{1+\Ga\over1-\Ga}\bigg)\cr
b_2&={12\over1-\Ga^2}\cr
\si_1\Ga^2&=b_0-2,\cr}
\eql{lowc2}$$
where $\Ga^2=\Ga^i\Ga_i$. The $\al_1$ and $\al_2$ in (\peq{lowc}) are
the standard numbers for the volume part.

Unfortunately
in this letter we will be able to show only the unsmeared results (\ie $f=1$),
although it is possible to extend the method of [\pref{BKD}] to cover the 
general situation (Dowker and Kirsten, in preparation).

As the simplest case, we take $V$ to be a $U(1)$ bundle
(a complex scalar) and the manifold $\man$ to be the 4-ball so
that $\pa\man$ is the 3-sphere. $\Ga^i$ is then just a real 
vector field on $\rS^3$. Nonsingular vector fields exist on all odd spheres of
course.

The condition (\peq{gbc}) can be rearranged
$$
{\cal B}=\nabla_N-\big(S-{i\over2}(\widehat\nabla_i\Ga^i)\big)+
i\Ga^i\widehat\nabla_i
\eql{gbc2}$$
Since $V$ is a scalar its covariant derivative is the ordinary one.

In order to apply the condition (\peq{gbc2}) we must fix on a simple 
$\Ga^i$. It is always advantageous to go to a local frame -- say a
right invariant one -- and introduce anholonomic coordinates through a set of
dreibeine,
$$
\pa_i=A_i^a\, X_a
$$
$$
\Ga^i=A^i_a\,\Ga^a,
\eql{comps}$$
where the $-iX_a$ are {\it right} angular momentum generators. We thus 
replace (\peq{gbc2}) by
$$
{\cal B}=\pa_r-S'+i\Ga^a X_a,
\eql{gbc3}$$
where $S'$ is the bracketed term in (\peq{gbc2}).

The condition (\peq{bcss}) must now be implemented on the modes which here 
take the form of a {\it sum} of boundary modes,
$$
\Phi(k)=\sum_\al C^\al(k){J_{\nu(\al)}(k r)\over r^{(d-1)/2}}\,Y_\al(\Om) 
\eql{modes2}$$
where the harmonics on the $d$-dimensional boundary satisfy
$$
\De_{\pa\man} Y(\Om)=-\la^2 Y(\Om)
$$
and 
$$
\nu^2= \la^2+(d-1)^2/4  . 
\eql{nus}$$
In the present special case $d=3$ and the surface harmonics are, 
conveniently, representation matrices of $SU(2)$,
$$
Y_\al(\Om)=\cau^{(L)}_{mn}(\Om);\quad \al=(L,m,n),
$$
where $\cau$ stands for the normalised eigenfunctions and $\la^2=4L(L+1)$ with
$L=0,1/2,1,\ldots$ so that from (\peq{nus}), $\nu=2L+1$.

The eigenmodes (\peq{modes2}) are finite at the origin and have nonzero 
eigenvalues $-k^2$.
There should be another label on $\Phi$ and $C^\al(k)$ to take over the role
of $\al$ which would come from solving some secular determinant. We shall not 
need it. 

In detail, (\peq{bcss}) is
$$
\sum_{L'm'n'} C^{(L'm'n')}\bigg(\cau^{(L')}_{m'n'}\big(\pa_r-S')
{J_{\nu(L')}(k r)\over r}\bigg|_{r=1}+
J_{\nu(L')}(k)i\Ga^a X_a\cau^{(L')}_{m'n'}\bigg) =0.
\eql{bcs2} $$

Now we use the right action 
$$
X_a\cau^{(L)}=2i\,\cau^{(L)}J_a,
\eql{ract}$$
expressed in matrix form, where the $J_a$ are the standard spin-$L$ angular
momentum matrices (the 2 is a normalisation factor)
and simplify by taking the $\Ga^a$ to be constants. By a property of the 
right-invariant dreibeine, $\Ga^i$ then has vanishing covariant divergence. 
This means $S=S'$ and also that the integrated flux across the boundary,
$
{i\over2}\int_{\pa\man}\big(V^*\nabla_NV-V\nabla_NV^*\big),
$
vanishes.

Multiplying (\peq{bcs2}) by $\cau^{(L)*}_{mn}$ and integrating over $\rS^3$ 
we get, using orthogonality,
$$
C^{(Lmn)}\big(\pa_r-S\big){J_{\nu(L)}(k r)\over r}
\bigg|_{r=1}- 2J_{\nu(L)}(k)\,\sum_{n'}C^{(Lmn')}{\bGa.\bf J}_{nn'}(L)=0.
\eql{bcs7}$$

As expected, we can drop the left $m$ labelling. Furthermore, if we make the 
special, diagonal choice,
$$
\bGa.{\bf J}= \Ga^0 J_z,
\eql{choice}$$
the sum in (\peq{bcs7}) reduces to one term ($n'=n$) and the $C$'s cancel to 
give 
$$
\big(\pa_r-S\big){J_{\nu(L)}(k r)\over r}
\bigg|_{r=1}- 2J_{\nu(L)}(k)\,\Ga^0\,n=0.
\eql{bcs8}$$

We see that the eigenvalues, $k^2$, depend on $n$ through this {\it 
generalised Robin condition} but we have 
obtained a condition to which our earlier methods can be applied, with
some modifications which will now be outlined.

The general technique has been described in sufficient detail in [\pref{BGKE}]
so that only the novelties arising in the present calculation need be 
indicated. 

The idea is to compute the coefficients from the analytical properties 
of the relevant \zf\ which can be found by progressive continuation
of an expression that incorporates the 
eigenvalue condition (\peq{bcs8}) through a contour integral. 

The method involves taking the logarithm of (\peq{bcs8}) and employing
the asymptotic behaviour of the Bessel functions in order to perform the 
continuation systematically, thereby gradually revealing the \zf's pole 
structure. The heat-kernel coefficients are related to the residues by a 
standard formula.

The condition (\peq{bcs8}) is slightly rewritten as
$$
kJ'_\nu(k)+(u+2gn)J_\nu(k)=0,
\eql{eigc}$$
where we have set $u=-S-1$ and $\Ga_0=-g$ for notational reasons.
Also $\nu=2L+1$ and, for each $L$, $n$ runs from $-L$ to 
$L$. It is the appearance of the $n$ in the last term that causes the
added complications.

Shifting the contour to the imaginary axis, the \zf\ reads
$$
\zeta (s) = {\sin \pi s\over \pi} \sum_{Lmn} 
\int_0^{\infty} dz \,\, (z\nu)^{-2s} {\partial \over \partial z} 
\log z^{-\nu}\bigg[ z\nu I_{\nu} ' (z\nu ) + (u+2gn) I_{\nu} (z\nu ) \bigg].
\eql{zeta}
$$
As shown in detail in [\pref{BKD}] the heat-kernel coefficients are  
determined solely by the asymptotic contributions of the Bessel functions 
as $\nu\to\infty$, but now more care is 
needed since terms like $n/\nu$ have to be counted as of order $\nu^0$. 

Applying this technique, one encounters the expression
$$\eqalign{
\log\bigg\{1+\bigg(1+\frac{2gn}{\nu} t \bigg)^{-1}
\bigg[\sum_{k=1}^{\infty} \frac{v_k (t) }{\nu^k}  +\frac{ut}{\nu}
&+\bigg( \frac{u+2gn}{\nu}\bigg)t\sum_{k=1}^{\infty} \frac{u_k(t)} {\nu^k}
\bigg]\bigg\}\cr
&=\sum_{j=1}^{\infty} {T_j (u,g,t)\over \nu^j} \cr}
\eql{asym}
$$
whereby the $T_j$ are defined and $t=1/\sqrt{1+z^2}$. For the Olver 
polynomials, $u_k$ and $v_k$, see \eg [\pref{AandS}].

Asymptotically one finds
$$
\zeta (s) = A_{-1} (s) +A_0 (s) +A_+ (s) +\sum_{j=1}^{\infty} A_j (s), 
\eql{zetasym}
$$
where $A_{-1} (s)$ and $A_0 (s)$ are the same as in Robin 
boundary conditions (see [\pref{BKD}]). The new quantities are 
$$
A_+ (s) = {\sin \pi s\over \pi} \sum_{Lmn}
\int_0^{\infty} dz \,\, (z\nu)^{-2s} {\partial \over \partial z}
\log\left( 1+{2gnt\over\nu} \right) ,
\eql{aplus}
$$
and 
$$A_j (s)  = {\sin \pi s\over \pi} \sum_{Lmn}
\int_0^{\infty} dz \,\, (z\nu)^{-2s} {\partial \over \partial z}
{T_j (u,g,t)\over \nu^j} .
\eql{aj}
$$

In order to proceed it is convenient to express $T_j$ as the finite sum
$$
T_j = \sum_{a,b,c} f_{a,b,c}^{(j)} {
 \de^c t^a \over \left( 1+\de t\right)^b },
\eql{tj}
$$
with $\delta = 2gn/\nu$.
The $f_{a,b,c}^{(j)} $ are easily determined via an algebraic 
computer programme. 

The next steps are to perform the $z$-integrations by the identity, 
$$\int_0^{\infty} dz \,\, z^{-2s} {zt^x\over (1+\delta t)^y } 
= \frac 1 2 {\Gamma (1-s) \over \Gamma (y) } \sum_{k=0}^{\infty}
(-1)^k {\Gamma (y+k) \Gamma (s-1+(x+k)/2) \over k! \Gamma ((x+k)/2) }
\delta ^k , \eql{zint}
$$
and then do the angular momentum summation, 
$$\sum_{Lmn} = \sum_{L=0,1/2,1,...}^{\infty} (2L+1) \sum_{n=-L}^L .\eql{angsum}
$$
The sum over $n$ may be treated in terms of Bernoulli numbers
 while that over $L$ produces Hurwitz zeta functions. In this 
way, all terms contributing to a heat-kernel coefficient can be 
determined systematically. 

In addition to the coefficients 
$a_0,a_{1/2}$, we find 
$$
a_1 = |S^3|\left( {2 \over 1-g^2 } S + 1 +{2 \over 1-g^2} -
{1\over g}\log{1+g\over1-g}\right)  
\eql{ay1}$$
where $|S^3|$ is the sphere volume.
A comparison of this expression with the general one, (\peq{lowc}), evaluated 
on the 4-ball in the notation  of [\pref{AandE}], {\it viz}
$$
a_1 = \frac 1 6 |S^3|
\left(3b_0 +Sb_2 +g^2 \sigma_1\right),
$$
yields
$$
b_2 = {12\over 1-g^2}\quad
{\rm and}\quad
\frac 1 2 b_0 +\frac 1 6 g^2 \sigma_1 = 1+{2\over 1-g^2} -{1\over g}
\log{1+g\over1-g}. 
$$
Together with the two equations from the conformal properties [\pref{AandE}],
$$
b_0-b_1-{1\over2}b_2+1=0\quad
{\rm and}\quad
b_0-2b_1-b_2+4-\si_1\Ga^2=0,
$$
one has three equations for three unknowns, the solution of which 
gives the known answer, (\peq{lowc2}) [\pref{McandO}]. 

Going beyond these results, we find the next two coefficients on the 4-ball,
$$\eqalign{
{a_{3/2} \over 8 \sqrt{\pi} |S^3|} =&{S \over 8g^2} \bigg( 1-
{1 \over \sqrt{1-g^2} } \bigg) \cr 
&+{1 \over (1-g^2)^{3/2} }\left(\frac 3 8 -\frac 1 {48g^4} -\frac 5 {32 g^2}
-\frac{11 g^2} {96} +\frac S 4 +\frac{S^2} 8 \right) \cr 
&+{1\over 1-g^2} \left( -\frac{107} {512} +\frac 1 {48g^4} +
\frac 1 {6g^2} +\frac{11g^2} {512} \right)\cr}
\eql{a32}$$
and
$$
{a_2 \over 8 |S^3|}
={29-28g^2-g^4+90S-30g^2S+90S^2+30S^3 \over 180 (g^2-1)^2 }.
\eql{a2} 
$$
It may be checked that the limit $g\to 0$ agrees with known results for
Robin conditions.

Avramidi and Esposito [\pref{AandE}] have given general forms for $a_{3/2}$ and
$a_2$ constructed from the allowed geometric objects subject to the restriction
that the boundary covariant derivative of $\Ga^i$ vanishes. 
Unfortunately this is not so in our case,
but Avramidi and Esposito's expression for $a_{3/2}$ can be augmented
to allow for a nonzero covariant derivative without too much effort.
For length considerations, this, and the comparison with our result, will be
carried out elsewhere. 

As is apparent from (\peq{lowc2}), something odd happens whenever any 
eigenvalue of $\Ga^2$ is greater than, or equal to, one. In fact our
specific expressions (\peq{ay1}) -- (\peq{a2}) exhibit 
branch points and poles at $g^2=1$. These singularities can be attributed to a 
loss of ellipticity in
the form of a breakdown of the Lopatinski condition, which, when 
satisfied, guarantees that the system has a unique solution. 
(See \eg [\pref{EandS,Gilkey1,krantz,treves}].)

In order to describe this condition, we note first that the leading symbol
of the Laplacian on $\pa\man$, $=\rS^3$, is $-g^{ij}\xi_i\xi_j=-\xi^a\xi_a
=-\bxi.\bxi$
in terms of anholonomic coordinates. Similarly the leading symbol of the
boundary condition (\peq{gbc2}) is $\pa_r+\bGa.\bxi$ and the classic 
Lopatinski condition requires that the set of equations
$$
\big(-\pa_r^2+\bxi^2\big)f(r)=0;\quad f(r)\to0\quad r\to\infty
\eql{L1}$$
and
$$
\big(-\pa_r-\bGa.\bxi\big)f(r)\big|_{\pa\man}=h(\bxi)
\eql{L2}$$
should have a unique solution for any $h(\bxi)$, for $|\bxi|\ne0$.

The relevant solution of the extended collar equation, (\peq{L1}), is
$$
f(r)=w(\bxi)\exp\big(-|\bxi|r\big)
$$
so that (\peq{L2}) reads
$$
\big(|\bxi|-\bGa.\bxi\big)w(\bxi)=|\bxi|\big(1-g\cos\th\big)w(\bxi)=h(\bxi)
$$
and there is a clear breakdown of invertibility when $g\ge1$ so that the
system is then not well-posed. $w(\bxi)$ is undetermined on a conical 
hypersurface in anholonomic coordinate space. (The argument extends to the 
case when the $\Ga^a$ are not constant and also to a general boundary.)

This situation is occasioned partly by the 
reality conditions imposed on the quantities occurring in the boundary 
condition (\peq{gbc}). In contrast, for the {\it Euclidean} string 
interacting with an electromagnetic field, [\pref{ACNY}],
these conditions produce no non-ellipticity, the 
beta-function, for example, being regular. However the operator seems not to
be self-adjoint.

In conclusion we have derived the scalar heat-kernel coefficients up to 
$a_2$ on the
four-ball for oblique boundary conditions of the simplest kind. 
An important extension would be that to arbitrary dimensions through higher 
spheres or tori, and this is under investigation.

A case of `practical' interest is quantum gravity. Here however the vector
bundle is that of symmetric tensors and the $\Ga^i$ are matrices composed 
from geometrical objects. These matrices
are such that $\Ga^2$ does not commute with $\Ga^i$, considerably 
complicating the  construction of the general forms [\pref{AandE}]. The 
technique of special case evaluation could be of assistance in this problem. 
It remains to be proved that the system is elliptic although one expects
that it should be.

In our opinion, the complicated character of the higher coefficients makes
their explicit general forms uninviting and it seems more likely that their
evaluation in particular cases will prove more valuable. However, the general
form would be needed if one were interested in evaluating the effective 
action from conformal variation. In four dimensions, this seems prohibitively
awkward. The two-dimensional case was done in [\pref{McandO}].
\section{\bf Acknowledgments}
We wish to thank Ivan Avramidi for assistance, in particular for an important 
clarification of the meaning of the eigenvalue condition.

This investigation has been partly supported by the DFG under contract number
BO1112/4-1.

\section{\bf References}
\vskip 5truept
\begin{putreferences}
\ref{APS}{Atiyah,M.F., V.K.Patodi and I.M.Singer: Spectral asymmetry and 
Riemannian geometry \mpcps{77}{75}{43}.}
\ref{AandT}{Awada,M.A. and D.J.Toms: Induced gravitational and gauge-field 
actions from quantised matter fields in non-abelian Kaluza-Klein thory 
\np{245}{84}{161}.}
\ref{BandI}{Baacke,J. and Y.Igarishi: Casimir energy of confined massive 
quarks \prD{27}{83}{460}.}
\ref{Barnesa}{Barnes,E.W.: On the Theory of the multiple Gamma function 
{\it Trans. Camb. Phil. Soc.} {\bf 19} (1903) 374.}
\ref{Barnesb}{Barnes,E.W.: On the asymptotic expansion of integral 
functions of multiple linear sequence, {\it Trans. Camb. Phil. 
Soc.} {\bf 19} (1903) 426.}
\ref{Barv}{Barvinsky,A.O. Yu.A.Kamenshchik and I.P.Karmazin: One-loop 
quantum cosmology \aop {219}{92}{201}.}
\ref{BandM}{Beers,B.L. and Millman, R.S. :The spectra of the 
Laplace-Beltrami
operator on compact, semisimple Lie groups. \ajm{99}{1975}{801-807}.}
\ref{BandH}{Bender,C.M. and P.Hays: Zero point energy of fields in a 
confined volume \prD{14}{76}{2622}.}
\ref{BBG}{Bla\v zi\' c,N., Bokan,N. and Gilkey,P.B.: Spectral geometry of the 
form valued Laplacian for manifolds with boundary \ijpam{23}{92}{103-120}}
\ref{BEK}{Bordag,M., E.Elizalde and K.Kirsten: { Heat kernel 
coefficients of the Laplace operator on the D-dimensional ball}, 
\jmp{37}{96}{895}.}
\ref{BGKE}{Bordag,M., B.Geyer, K.Kirsten and E.Elizalde,: { Zeta function
determinant of the Laplace operator on the D-dimensional ball}, 
\cmp{179}{96}{215}.}
\ref{BKD}{Bordag,M., K.Kirsten,K. and Dowker,J.S.: Heat kernels and
functional determinants on the generalized cone \cmp{182}{96}{371}.}
\ref{Branson}{Branson,T.P.: Conformally covariant equations on differential
forms \cpde{7}{82}{393-431}.}
\ref{BandG2}{Branson,T.P. and Gilkey,P.B. {\it Comm. Partial Diff. Eqns.}
{\bf 15} (1990) 245.}
\ref{BGV}{Branson,T.P., P.B.Gilkey and D.V.Vassilevich {\it The Asymptotics
of the Laplacian on a manifold with boundary} II, hep-th/9504029.}
\ref{BCZ1}{Bytsenko,A.A, Cognola,G. and Zerbini, S. : Quantum fields in
hyperbolic space-times with finite spatial volume, hep-th/9605209.}
\ref{BCZ2}{Bytsenko,A.A, Cognola,G. and Zerbini, S. : Determinant of 
Laplacian on a non-compact 3-dimensional hyperbolic manifold with finite
volume, hep-th /9608089.}
\ref{CandH2}{Camporesi,R. and Higuchi, A.: Plancherel measure for $p$-forms
in real hyperbolic space, \jgp{15}{94}{57-94}.} 
\ref{CandH}{Camporesi,R. and A.Higuchi {\it On the eigenfunctions of the 
Dirac operator on spheres and real hyperbolic spaces}, gr-qc/9505009.}
\ref{ChandD}{Chang, Peter and J.S.Dowker :Vacuum energy on orbifold factors
of spheres, \np{395}{93}{407}.}
\ref{cheeg1}{Cheeger, J.: Spectral Geometry of Singular Riemannian Spaces.
\jdg {18}{83}{575}.}
\ref{cheeg2}{Cheeger,J.: Hodge theory of complex cones {\it Ast\'erisque} 
{\bf 101-102}(1983) 118-134}
\ref{Chou}{Chou,A.W.: The Dirac operator on spaces with conical 
singularities and positive scalar curvature, \tams{289}{85}{1-40}.}
\ref{CandT}{Copeland,E. and Toms,D.J.: Quantized antisymmetric 
tensor fields and self-consistent dimensional reduction 
in higher-dimensional spacetimes, \break\np{255}{85}{201}}
\ref{DandH}{D'Eath,P.D. and J.J.Halliwell: Fermions in quantum cosmology 
\prD{35}{87}{1100}.}
\ref{cheeg3}{Cheeger,J.:Analytic torsion and the heat equation. \aom{109}
{79}{259-322}.}
\ref{DandE}{D'Eath,P.D. and G.V.M.Esposito: Local boundary conditions for 
Dirac operator and one-loop quantum cosmology \prD{43}{91}{3234}.}
\ref{DandE2}{D'Eath,P.D. and G.V.M.Esposito: Spectral boundary conditions 
in one-loop quantum cosmology \prD{44}{91}{1713}.}
\ref{Dow1}{Dowker,J.S.: Effective action on spherical domains, \cmp{162}{94}
{633}.}
\ref{Dow8}{Dowker,J.S. {\it Robin conditions on the Euclidean ball} 
MUTP/95/7; hep-th\break/9506042. {\it Class. Quant.Grav.} to be published.}
\ref{Dow9}{Dowker,J.S. {\it Oddball determinants} MUTP/95/12; 
hep-th/9507096.}
\ref{Dow10}{Dowker,J.S. {\it Spin on the 4-ball}, 
hep-th/9508082, {\it Phys. Lett. B}, to be published.}
\ref{DandA2}{Dowker,J.S. and J.S.Apps, {\it Functional determinants on 
certain domains}. To appear in the Proceedings of the 6th Moscow Quantum 
Gravity Seminar held in Moscow, June 1995; hep-th/9506204.}
\ref{DABK}{Dowker,J.S., Apps,J.S., Bordag,M. and Kirsten,K.: Spectral 
invariants for the Dirac equation with various boundary conditions 
{\it Class. Quant.Grav.} to be published, hep-th/9511060.}
\ref{EandR}{E.Elizalde and A.Romeo : An integral involving the
generalized zeta function, {\it International J. of Math. and 
Phys.} {\bf13} (1994) 453.}
\ref{ELV2}{Elizalde, E., Lygren, M. and Vassilevich, D.V. : Zeta function 
for the laplace operator acting on forms in a ball with gauge boundary 
conditions. hep-th/9605026}
\ref{ELV1}{Elizalde, E., Lygren, M. and Vassilevich, D.V. : Antisymmetric
tensor fields on spheres: functional determinants and non-local
counterterms, \jmp{}{96}{} to appear. hep-th/ 9602113}
\ref{Kam2}{Esposito,G., A.Y.Kamenshchik, I.V.Mishakov and G.Pollifrone: 
Gravitons in one-loop quantum cosmology \prD{50}{94}{6329}; 
\prD{52}{95}{3457}.}
\ref{Erdelyi}{A.Erdelyi,W.Magnus,F.Oberhettinger and F.G.Tricomi {\it
Higher Transcendental Functions} Vol.I McGraw-Hill, New York, 1953.}
\ref{Esposito}{Esposito,G.: { Quantum Gravity, Quantum Cosmology and 
Lorentzian Geometries}, Lecture Notes in Physics, Monographs, Vol. m12, 
Springer-Verlag, Berlin 1994.}
\ref{Esposito2}{Esposito,G. {\it Nonlocal properties in Euclidean Quantum
Gravity}. To appear in Proceedings of 3rd. Workshop on Quantum Field Theory
under the Influence of External Conditions, Leipzig, September 1995; 
gr-qc/9508056.}
\ref{EKMP}{Esposito G, Kamenshchik Yu A, Mishakov I V and Pollifrone G.:
One-loop Amplitudes in Euclidean quantum gravity.
\prd {52}{96}{3457}.}
\ref{ETP}{Esposito,G., H.A.Morales-T\'ecotl and L.O.Pimentel {\it Essential
self-adjointness in one-loop quantum cosmology}, gr-qc/9510020.}
\ref{FORW}{Forgacs,P., L.O'Raifeartaigh and A.Wipf: Scattering theory, U(1) 
anomaly and index theorems for compact and non-compact manifolds 
\np{293}{87}{559}.}
\ref{GandM}{Gallot S. and Meyer,D. : Op\'erateur de coubure et Laplacian
des formes diff\'eren-\break tielles d'une vari\'et\'e riemannienne 
\jmpa{54}{1975}
{289}.}
\ref{Gilkey1}{Gilkey, P.B, Invariance theory, the heat equation and the
Atiyah-Singer index theorem, 2nd. Edn., CRC Press, Boca Raton 1995.}
\ref{Gilkey2}{Gilkey,P.B.:On the index of geometric operators for 
Riemannian manifolds with boundary \aim{102}{93}{129}.}
\ref{Gilkey3}{Gilkey,P.B.: The boundary integrand in the formula for the 
signature and Euler characteristic of a manifold with boundary 
\aim{15}{75}{334}.}
\ref{Grubb}{Grubb,G. {\it Comm. Partial Diff. Eqns.} {\bf 17} (1992) 
2031.}
\ref{GandS1}{Grubb,G. and R.T.Seeley \cras{317}{1993}{1124}; \invm{121}{95}
{481}.}
\ref{GandS}{G\"unther,P. and Schimming,R.:Curvature and spectrum of compact
Riemannian manifolds, \jdg{12}{77}{599-618}.}
\ref{IandT}{Ikeda,A. and Taniguchi,Y.:Spectra and eigenforms of the 
Laplacian
on $S^n$ and $P^n(C)$. \ojm{15}{1978}{515-546}.}
\ref{IandK}{Iwasaki,I. and Katase,K. :On the spectra of Laplace operator
on $\La^*(S^n)$ \pja{55}{79}{141}.}
\ref{JandK}{Jaroszewicz,T. and P.S.Kurzepa: Polyakov spin factors and 
Laplacians on homogeneous spaces \aop{213}{92}{135}.}
\ref{Kam}{Kamenshchik,Yu.A. and I.V.Mishakov: Fermions in one-loop quantum 
cosmology \prD{47}{93}{1380}.}
\ref{KandM}{Kamenshchik,Yu.A. and I.V.Mishakov: Zeta function technique for
quantum cosmology {\it Int. J. Mod. Phys.} {\bf A7} (1992) 3265.}
\ref{KandC}{Kirsten,K. and Cognola.G,: { Heat-kernel coefficients and 
functional determinants for higher spin fields on the ball} \cqg{13}{96}
{633-644}.}
\ref{Levitin}{Levitin,M.: { Dirichlet and Neumann invariants for Euclidean
balls}, {\it Diff. Geom. and its Appl.}, to be published.}
\ref{Luck}{Luckock,H.C.: Mixed boundary conditions in quantum field theory 
\jmp{32}{91}{1755}.}
\ref{MandL}{Luckock,H.C. and Moss,I.G,: The quantum geometry of random 
surfaces and spinning strings \cqg{6}{89}{1993}.}
\ref{Ma}{Ma,Z.Q.: Axial anomaly and index theorem for a two-dimensional 
disc 
with boundary \jpa{19}{86}{L317}.}
\ref{Mcav}{McAvity,D.M.: Heat-kernel asymptotics for mixed boundary 
conditions \cqg{9}{92}{1983}.}
\ref{MandV}{Marachevsky,V.N. and D.V.Vassilevich {\it Diffeomorphism
invariant eigenvalue \break problem for metric perturbations in a bounded 
region}, SPbU-IP-95, \break gr-qc/9509051.}
\ref{Milton}{Milton,K.A.: Zero point energy of confined fermions 
\prD{22}{80}{1444}.}
\ref{MandS}{Mishchenko,A.V. and Yu.A.Sitenko: Spectral boundary conditions 
and index theorem for two-dimensional manifolds with boundary 
\aop{218}{92}{199}.}
\ref{Moss}{Moss,I.G.: Boundary terms in the heat-kernel expansion 
\cqg{6}{89}{759}.}
\ref{MandP}{Moss,I.G. and S.J.Poletti: Conformal anomaly on an Einstein space 
with boundary \pl{B333}{94}{326}.}
\ref{MandP2}{Moss,I.G. and S.J.Poletti \np{341}{90}{155}.}
\ref{NandOC}{Nash, C. and O'Connor,D.J.: Determinants of Laplacians, the 
Ray-Singer torsion on lens spaces and the Riemann zeta function 
\jmp{36}{95}{1462}.}
\ref{NandS}{Niemi,A.J. and G.W.Semenoff: Index theorem on open infinite 
manifolds \np {269}{86}{131}.}
\ref{NandT}{Ninomiya,M. and C.I.Tan: Axial anomaly and index thorem for 
manifolds with boundary \np{245}{85}{199}.}
\ref{norlund2}{N\"orlund~N. E.:M\'emoire sur les polynomes de Bernoulli.
\am {4}{21} {1922}.}
\ref{Poletti}{Poletti,S.J. \pl{B249}{90}{355}.}
\ref{RandT}{Russell,I.H. and Toms D.J.: Vacuum energy for massive forms 
in $R^m\times S^N$, \cqg{4}{86}{1357}.}
\ref{RandS}{R\"omer,H. and P.B.Schroer \pl{21}{77}{182}.}
\ref{Trautman}{Trautman,A.: Spinors and Dirac operators on hypersurfaces 
\jmp{33}{92}{4011}.}
\ref{Vass}{Vassilevich,D.V.{Vector fields on a disk with mixed 
boundary conditions} gr-qc /9404052.}
\ref{Voros}{Voros,A.:
Spectral functions, special functions and the Selberg zeta function.
\cmp{110}{87}439.}
\ref{Ray}{Ray,D.B.: Reidemeister torsion and the Laplacian on lens
spaces \aim{4}{70}{109}.}
\ref{McandO}{McAvity,D.M. and Osborn,H. Asymptotic expansion of the heat kernel
for generalised boundary conditions \cqg{8}{91}{1445}.}
\ref{AandE}{Avramidi,I. and Esposito,G. Heat kernel asymptotics with 
generalised boundary conditions, hep-th/9701018.}
\ref{MandS}{Moss,I.G. and Silva P.J., Invariant boundary conditions for
gauge theories gr-qc/9610023.}
\ref{barv}{Barvinsky,A.O.\pl{195B}{87}{344}.}
\ref{krantz}{Krantz,S.G. Partial Differential Equations and Complex
Analysis (CRC Press, Boca Raton, 1992).}
\ref{treves}{Treves,F. Introduction to Pseudodifferential and Fourier Integral
Operators,\break Vol.1, (Plenum Press,New York,1980).}
\ref{EandS}{Egorov,Yu.V. and Shubin,M.A. Partial Differential Equations
(Springer-Verlag, Berlin,1991).}
\ref{AandS}{Abramowitz,M. and Stegun,I.A. Handbook of Mathematical Functions 
(Dover, New York, 1972).}
\ref{ACNY}{Abouelsaood,A., Callan,C.G., Nappi,C.R. and Yost,S.A.\np{280}{87}
{599}.}
\ref{BGKE}{Bordag,M., B.Geyer, K.Kirsten and E.Elizalde, { Zeta function
determinant of the Laplace operator on the D-dimensional ball}, 
\cmp{179}{96}{215}.}

\end{putreferences}
\bye